\documentclass[sigconf]{acmart}

\usepackage{array}
\usepackage{multirow}
\usepackage{makecell}
\usepackage{array}
\usepackage{makecell}
\usepackage{url}
\usepackage{bbding}
\usepackage{pifont}
\usepackage{wasysym}
\AtBeginDocument{%
  \providecommand\BibTeX{{%
    \normalfont B\kern-0.5em{\scshape i\kern-0.25em b}\kern-0.8em\TeX}}}

\copyrightyear{2023}
\acmYear{2023}
\begin{document}

%%
%% The "title" command has an optional parameter,
%% allowing the author to define a "short title" to be used in page headers.
\title[Quaternary Studies in Software Engineering]{Anachronic Tertiary Studies in Software Engineering: An Exploratory Quaternary Study}

%%
%% The "author" command and its associated commands are used to define
%% the authors and their affiliations.
%% Of note is the shared affiliation of the first two authors, and the
%% "authornote" and "authornotemark" commands
%% used to denote shared contribution to the research.
%%\authornote{Both authors contributed equally to this research.}

%\orcid{1234-5678-9012}
%\authornotemark[1]

%\author{\mbox{Hidden for blind review purposes}}
%\affiliation{%
  %\institution{Institution}
  %\city{City}
  %\state{State}
  %\country{Country}
 % }
%\email{Email}%

%\begin{comment}
\author{\mbox{Valdemar Vicente Graciano Neto}}
\affiliation{%
  \institution{Federal University of Goiás}
  \city{Goiânia}
  \state{Goiás}
  \country{Brazil}
  }
\email{valdemarneto@ufg.br}

\author{\mbox{Célia Laís Rodrigues}}
\affiliation{%
  \institution{Federal University of Goiás}
  \city{Goiânia}
  \state{Goiás}
  \country{Brazil}
  }
\email{celialais@egresso.ufg.br}
  
\author{Fernando Kenji Kamei}
\affiliation{%
  \institution{Federal Institute of Alagoas}
  \city{Maceió}
  \state{Alagoas}
  \country{Brazil}
  }
\email{fernando.kenji@ifal.edu.br}

\author{Juliano Lopes de Oliveira}
\affiliation{%
  \institution{Federal University of Goiás}
  \city{Goiânia}
  \state{Goiás}
  \country{Brazil}
  }
\email{juliano@inf.ufg.br}

\author{Eliomar Araújo de Lima}
\affiliation{%
  \institution{Federal University of Goiás}
  \city{Goiânia}
  \state{Goiás}
  \country{Brazil}
  }
\email{eliomar.lima@ufg.br}

\author{Mohamad Kassab}
\affiliation{%
  \institution{Penn State University}
  \city{Malvern}
  \state{Pennsylvania}
  \country{United States}
  }
\email{muk36@psu.edu}

\author{Roberto Oliveira}
\affiliation{%
  \institution{State University of Goiás}
  \city{Posse}
  \state{Goiás}
  \country{Brazil}
  }
\email{roberto.oliveira@ueg.br}
%\end{comment}

%\renewcommand{\shortauthors}{Graciano Neto et al.}
\renewcommand{\shortauthors}{Hidden for blind review}

%%
%% By default, the full list of authors will be used in the page
%% headers. Often, this list is too long, and will overlap
%% other information printed in the page headers. This command allows
%% the author to define a more concise list
%% of authors' names for this purpose.
%\renewcommand{\shortauthors}{Trovato and Tobin, et al.}

%%
%% The abstract is a short summary of the work to be presented in the
%% article.
\begin{abstract}
%Systematic literature reviews (including mapping, secondary and tertiary studies) are attributed the predicate of representing the state of the art in their area of investigation. However,in cases such as tertiary studies, the publication of other primary studies after the publication of the secondary studies included can quickly outdate the tertiary study, or even the active publication of new primary and secondary studies that takes place after the publication of a tertiary study. As a consequence, when using that study as a reference for certain branches of knowledge, imprecision may be incurred, both from the point of view of its subareas and of new methodologies, languages, tools and others. The main contribution of this paper is to communicate emerging results of an analysis of tertiary studies, against the background of tertiary studies in software engineering, that is, a literature review that included and analyzed tertiary studies exclusively. 34 tertiary studies published between 2009 and 2021 were included. Results show that just over 60\% of the analyzed studies may have some degree of anachronism due to the publication of new primary and secondary studies in the year following the publication of the tertiary study analyzed.

Systematic literature reviews tentatively%, encompassing mapping, secondary, and tertiary studies, are widely regarded as representing 
describe the state of the art in a given research area. However, the continuous publication of new primary and secondary studies following the release of a tertiary study can make the communication of results not integrally representative in regards to the advances achieved by that time. Consequently, using such a study as a reference within specific bodies of knowledge may introduce imprecision, both in terms of its subareas and with respect to new methodologies, languages, and tools. Thus, a review of tertiary studies (what could be understood as a quaternary study) could contribute to show the representativeness of the reported findings in comparison to the state of the art and also to compile a set of perceptions that could not be previously achieved. In that direction, the main contribution of this paper is presenting the findings from an analysis of 34 software engineering tertiary studies published between 2009 and 2021. 
The results indicate that over 60\% of the studies demonstrate varying degrees of anachronism due to the publication of primary and secondary studies following the publication of the tertiary study or even due to a time elapse between its conduction and its publication.
\end{abstract}

%%
%% The code below is generated by the tool at http://dl.acm.org/ccs.cfm.
%% Please copy and paste the code instead of the example below.
%%
% \begin{CCSXML}
% <ccs2012>
%  <concept>
%   <concept_id>10010520.10010553.10010562</concept_id>
%   <concept_desc>Computer systems organization~Embedded systems</concept_desc>
%   <concept_significance>500</concept_significance>
%  </concept>
%  <concept>
%   <concept_id>10010520.10010575.10010755</concept_id>
%   <concept_desc>Computer systems organization~Redundancy</concept_desc>
%   <concept_significance>300</concept_significance>
%  </concept>
%  <concept>
%   <concept_id>10010520.10010553.10010554</concept_id>
%   <concept_desc>Computer systems organization~Robotics</concept_desc>
%   <concept_significance>100</concept_significance>
%  </concept>
%  <concept>
%   <concept_id>10003033.10003083.10003095</concept_id>
%   <concept_desc>Networks~Network reliability</concept_desc>
%   <concept_significance>100</concept_significance>
%  </concept>
% </ccs2012>
% \end{CCSXML}

% \ccsdesc[500]{Computer systems organization~Embedded systems}
% \ccsdesc[300]{Computer systems organization~Redundancy}
% \ccsdesc{Computer systems organization~Robotics}
% \ccsdesc[100]{Networks~Network reliability}

%%
%% Keywords. The author(s) should pick words that accurately describe
%% the work being presented. Separate the keywords with commas.

%% A "teaser" image appears between the author and affiliation
%% information and the body of the document, and typically spans the
%% page.
% \begin{teaserfigure}
%   \includegraphics[width=\textwidth]{sampleteaser}
%   \caption{Seattle Mariners at Spring Training, 2010.}
%   \Description{Enjoying the baseball game from the third-base
%   seats. Ichiro Suzuki preparing to bat.}
%   \label{fig:teaser}
% \end{teaserfigure}

%%
%% This command processes the author and affiliation and title
%% information and builds the first part of the formatted document.
\maketitle

\section{Introduction}

Systematic Literature Reviews (SLR) are roughly important for the science advance, particularly in Software Engineering. They use to consume a considerable effort, but also bring important insights for the state of the art. Tertiary Studies (TS\footnote{Henceforth, this acronym will be interchangeably used to express both singular and plural forms: Tertiary study and Tertiary studies.}) are a particular type of SLR that gather and consolidate knowledge from multiple secondary studies, potentially reflecting the state of the art on specific domains \cite{Kudo20}. %Nevertheless, akin to secondary studies, TS are susceptible to obsolescence due to the publication of novel studies that they do not cover.
For instance, Cadavid et al. (2020) \cite{Cadavid2020} is a insightful tertiary study that aggregates secondary studies concerning Systems-of-Systems (SoS) Architecting. The study was well-conducted and revealed us that, although the study included a secondary study published in 2015 \cite{Guessi2015} that maps the architectural description languages (ADL) for SoS, it did not incorporate a significant ADL developed explicitly for SoS, published in the subsequent year (2016) in a primary study \cite{Oquendo2016}. This finding does not invalidate or question the results communicated by the authors. However, it allows us to illustrate an important (and known) side effect of TS or of any SLR: a possible gap (and a temporal void) between the year of publication of literature review (2015, in that case) and possible advances (primary studies) that appear in the following years until its publication (2020, in that case). %This omission stems from the fact that the secondary study addressing the subject of architectural description of SoS had already become anachronic with regard to the scientific progress made in the intervening years. 
%Another case examined within this study pertains to the research conducted by Costa et al. (2020) \cite{Costa2020}, which presents a tertiary study on microservices. Subsequently, in 2021, a secondary study was published, focusing on the current state of implementation of microservices \cite{Nino2021}. Furthermore, in the same year, 16 primary studies addressing the same subject were also published, all of which were encompassed within the scope of this investigation. Consequently, the study published in 2020 has become anachronic by 2021, owing to the emergence of novel secondary and primary studies that surpass its coverage.

%Quando acadêmicos e práticos recorrem a estudos terciários, espera-se que ele represente o estado da arte e que se possa utilizá-lo como fonte de consulta, dado que foi sistematicamente conduzido, pode ser auditável e trata-se de resultados baseados em evidência. Entretanto, ainda que numa pesquisa ou desenvolvimento de produto de pesquisa e inovação um estudo terciário obviamente não será a única fonte de consulta, a desatualização acidental causada pela natureza do próprio estudo pode gerar prejuízos importantes, tais como a decisão por utilizar uma tecnologia ou método que não corresponde ao que há de mais recente ou avançado na área.
%The main problems that could arise from such a phenomenon relies on damages that could potentially be caused by such a situation.
The need to update SLRs is a well-discussed topic in the literature \cite{Mendes2020,wohlin2020guidelines,wohlin2022successful,mendes2019search,felizardo2018evaluating,garner2016and}. % given their systematic nature, verifiability, and reliance on evidence-based findings. 
In general, the current studies investigate when, how and why to update SLR at the level of a secondary study. We are not aware of initiatives that investigate what should justify the update of TS. While TS will not likely be the sole source of reference in an R\&D project (since Rapid Reviews are the approach often adopted for that purpose \cite{cartaxo2018role,cartaxo2020rapid}), the accidental obsolescence stemming from the inherent characteristics of such studies could lead to drawbacks. For instance, reliance on anachronic technology or methodologies that do not align with the most recent advances in the field could undermine decision-making.   

To enlighten this issue, the main contribution of this vision paper is to communicate \textbf{preliminary results} of a review of TS, what could be called as a quaternary study. Focusing on the analysis of TS exclusively, the \textbf{novelty of this research is threefold}:% (i) to examine the current state of TS within the field, and (ii) to gain insights into the fundamental attributes that may define a quaternary study. 
1) to shed light in a discussion about what a quaternary study in software engineering could be; 2) externalize a study that exclusively addresses TS and investigates which of them had primary and secondary studies published after their publication and were not necessarily included/considered; 3) reflect about what updating TS would mean and the differences between updating a secondary study and a TS, given that the current state of the art only deals with updating SLR (secondary studies). For that purpose, we conducted a quaternary study, that is, a systematic literature review that includes only TS in software engineering. From 206 studies initially  retrieved from two scientific databases, 34 studies were included and analyzed. Results reveal that 60\% (21 of 34) TS analyzed may have some degree of anachronism due to primary and secondary studies published (and not considered in the TS) after the publication of the analyzed studies.

This paper is structured as follows: Section \ref{sec:background} provides background and related work. Section \ref{sec:propsolution} presents the review protocol, details about the conduction, and results reporting. %that support us towards achieving the proposed solution and evaluating it accordingly.
Section \ref{sec:progs} discusses the advances so far and Section \ref{sec:finalremarks} brings final remarks.
\vspace{-0.3cm}

\section{Background}
\label{sec:background}

Evidence-Based Software Engineering (ESBE) is a research paradigm 
%introduced by Kitchenham, Dyba and Jorgensen \cite{Kitchenham2004}, 
rooted in the principles of evidence-based practice and systematic literature reviews. 
%commonly employed in the medical, education and economics domains. 
%Systematic Literature Review (SLR) is one of the most common types of secondary study, as is Systematic Mapping Study (SMS) \cite{ Kitchenham, KITCHENHAM2011, PETERSEN20151}. 
Primary studies are peer-reviewed empirical studies that aim to investigate specific research questions and to describe evidence-based practices. They can report results of controlled experiments, case studies, surveys and other primary sources of research data \cite{Kitchenham2007, Petticrew2005}. Secondary studies identify, analyze and evaluate primary studies that can answer research questions to systematically synthesize evidence about that domain \cite{Kitchenham2004}. 
TS are \textit{systematic reviews of secondary studies} that raise, analyze, catalog and synthesize research data and scientific evidences exclusively from secondary studies \cite{Kitchenham2007}. TS in Software Engineering started with the works of Kitchenham et al. \cite{KITCHENHAM2009,KITCHENHAM2010} and, according to Garousi and Mäntylä \cite{Garousi2016}, the number of secondary studies reviewed by a TS varies from 12 to 116. %A TS can be considered \textbf{anachronic} for two reasons: by the emergence of new studies that were not included; or by including older secondary studies that do not take into account advances reported in primary studies published in subsequent years.

Given the large number of SLR published over the years in Software Engineering (SE) and the relatively low number of updated SLR, many SLRs in SE are potentially anachronic, which can importantly affect the current understanding of the state-of-the-art in those SLRs’ research topics, as stated by Mendes et al. (2020) \cite{Mendes2020}. In the same study, the authors recall the definition of what can be considered \textbf{an update of a SLR}, extracted from the report of a panel with experts in that subject \cite{garner2016and}: \textit{an update of a SLR [is defined] as a new edition of a published SLR with changes that can include new data, new methods, or new analyses to the previous edition. An update asks a similar question with regard to  the  participants,  intervention,  comparisons,  and  outcomes  (PICO)  and  has similar objectives; thus, it has similar inclusion criteria. These inclusion criteria can  be  modified  in  the  light  of  developments  within  the  topic  area  with  new interventions, new standards, and new approaches. Updates will include a new search  for  potentially  relevant  studies  and  incorporate  any  eligible  studies  or data; and adjust the findings and conclusions as appropriate.}

In their study, Mendes et al. (2020) \cite{Mendes2020} included and analyzed studies under three categories: i) systematic review about when and how to update SLRs; ii) techniques to identify new evidence related to previously published SLRs; and iii) 
decision mechanisms (factors or decision tree) to decide whether an SLR needs updating. The authors algo gather factors that can affect the decision as to whether an SLR update is needed, and they are: a) The SLR’s topic is still relevant, and some preliminary searches 
suggest that there are new studies suitable for inclusion; b) The SLR’s topic is relatively new, and the original SLR included limited data. An update should be done if it is acknowledged that such an update would provide valuable additional information; c)  Large volumes of information have been published over a short timescale; and d) Large  influential  studies  were  published  and  may  affect  the  original  SLR’s conclusions. 

The authors also recall the  decision framework (third-party  decision  framework, known as 3PDF) proposed by Garner et al. \cite{garner2016and}, which include the following steps to analyze if an SLR should be updated: Step 1) Assessing how current/actual the SLR is by looking at its topic’s relevance  for  research  and  practice, including an analysis of the impact  on research and/or practice (using metrics such as citations via sites such as Google Scholar), and finally whether the SLR was carried out properly and using a sound methodology; Step  2)  identifying  if  there  are  any  new  methods  proposed  and/or  new  studies published after the SLR’s publication, with Step 2.b specifically analyzing whether new additional studies were found; and Step 3) assessing whether the adoption of new methods and/or new studies may affect the conclusions when compared to the conclusions from the original SLR, and/or the original SLR’s credibility. 

The main conclusions of the authors were (i) 14  of  the  20  SE  SLR  updates  did  not  need updating; and (ii) the main decision driver to whether an SLR should be updated or not was Step 1b - the SLR’s contribution to research and practice. We perceive that all the findings so far are related to SLR as secondary studies, but that the need to update TS have not been largely discussed.
\\
\noindent \textbf{Related Work.} During our investigation, we encountered a sole study that also undertook the examination of TS. Published in October 2021, the study conducted by Napoleao et al. \cite{Napoleao2021} aimed to survey and analyze TS with the objective of formulating a more suitable search string to effectively identify secondary studies within SR. Other prior studies also investigated the up-to-date state of literature reviews and how/when to update them \cite{felizardo:2016update,Mendes2020}. However, we are not aware of other studies that exclusively review  TS and/or investigate the nature of what could be considered as a quaternary study. 

\section{A Review of Tertiary Studies in Software Engineering}
\label{sec:propsolution}

For building upon the concept of updating a TS, we proposed the term \textit{\textbf{anachronic}}\footnote{Anachronism is the concept that refers to using the concepts and ideas of a given time to analyze the facts of another time. In our context, we use this concept to denote a possible chronological inconsistency, as in the case of TS that include secondary studies which may potentially required updating. Other terms for Anachronic include anachronical, anachronous and anachronistic.} to describe a TS that potentially need an update. In a first moment, we consider a TS can be considered \textbf{anachronic} for two likely reasons: by the emergence of new secondary studies of that topic that were not included; or by including older secondary studies that do not take into account advances reported in primary studies published in subsequent years. A TS, from our point of view, can be considered anachronistic when it analyzes a specific time frame, potentially neglecting relevant studies in that area, which may indicate it needs some further evidence to support its results or that some of its included secondary studies may need updating in light of
criteria defined by the state of the art on SLR updating.

This section describes the protocol designed and used to conduct our quaternary study. Our protocol followed the guidelines of Kitchenham and Charles \cite{Kitchenham2007}, structured in \textit{Planning}, \textit{Conduction}, and \textit{Reporting}, involving five researchers.
\vspace{-0.2cm}
\subsection{Step 1: Planning (Study Protocol)}
\label{sec:systematic_mapping_results}

As stated earlier, the primary \textbf{aim of this research} was twofold: (i) \textbf{to examine the current state of tertiary studies within the Software Engineering area}, and (ii) \textbf{to gain enhanced insights into the fundamental attributes that may define a quaternary study}. Research Questions (RQ) were elaborated to address the former, whilst an analysis and discussion of the results can help us to conjecture insights for the latter. The following research questions (RQ) were raised:
\\
\textbf{RQ1: \textit{What are the covered areas in tertiary studies?}} \textbf{Rationale:} This question aims to identify which areas and subareas have the highest volume of tertiary studies.\\
\textbf{RQ2:} \textbf{\textit{What is the year of the oldest and most recent secondary study analyzed by the included tertiary studies?}} \textbf{Rationale:} The year of publication allows us to investigate, identify and compare existing secondary and primary studies on the same area. \\
%\textbf{RQ3:} \textbf{\textit{Is there evidence that the research carried out in the tertiary study was limited by the lack of secondary studies in the addressed area?}}
%\textbf{Rationale:} The limitations and quality of the study reflect the number of secondary studies, if a certain area has a shortage of quality secondary studies.\\
%\textbf{RQ3: \textit{Are there TS in the same area published in years prior to the included TS?}} \textbf{Rationale:} By answering this question, we can identify whether that TS was already covered or updated.\\
\textbf{RQ3: \textit{Are there secondary studies published after the TS in the same area that might make it anachronic? If yes, what is the year of publication?}} \textbf{Rationale:} By answering this question, we aim to identify whether the TS has evidence that it may be anachronic due to the existence of secondary studies in the same area and identify the time difference between the publication of the TS analyzed and the secondary study found.\\
\textbf{RQ4: \textit{Are there primary studies published after the TS in the same area that might make it anachronic? If yes, what is the year of publication?}} \textbf{Rationale:} We aimed at discovering whether the included TS had evidence that might be anachronic due to the existence of primary studies in the same area.
\\
%To answer RQ3 and RQ4, other individual searches were conducted in the same databases. %In turn, to assess the quality of the selected studies, quality assessment questions (QQ) were defined. These questions were based on questions commonly used in studies of evidence-based literature reviews \cite{Teixeira2019, ALI2010, DYBA2008}.\\

%\textbf{QQ1:} Was a rationale presented in the study for its execution?

%\textbf{QQ2:} Did the study address prior work in the targeted area?

%\textbf{QQ3:} Was the context in which the research was conducted accordingly described?

%\textbf{QQ4:} Was the research method founded and justified?

%\textbf{QQ5:} Was the contribution generated by the study presented with sufficient data to justify it?

%\textbf{QQ6:} Were the limitations and credibility of the study discussed?

%\textbf{QQ7:} Has future work been presented based on the contributions of the study?\\
\vspace{-0.2cm}
\textbf{Search  Strategy.} The search strategy chosen was an automatic search performed in two databases (namely ACM Digital Library\footnote{\url{https://dl.acm.org/}} and IEEE Xplore\footnote{\url{https://ieeexplore.ieee.org/}}). According to Dyba \textit{et al.} \cite{petersen2015guidelines} and Kitchenham and Charters \cite{Kitchenham2007}, these publication databases are some of the most relevant sources in the Computer Science and Information Systems areas. Only these two databases were chosen due to the exploratory nature of this study.

We used the following search string: \texttt{``tertiary studies''}. %We did not use other terms in the search string for some reasons: 1) The chosen scientific databases mostly contain published research in computing and its derived areas, including Software Engineering, and 2) Tertiary studies, in the Computer Science area, are majorly conducted in Software Engineering research sub-branches. 
Then, any eventual tertiary study from other areas could be excluded during the selection step. To answer the research questions RQ3 and RQ4, searches were carried out in the same libraries where the tertiary studies were collected, IEEE Xplorer and ACM Digital Library. The search strings used to raise the secondary and primary studies were the same used in the tertiary study being analyzed in addition with the keywords presented by that TS.
\\
\textbf{Inclusion and Exclusion Criteria.} The Inclusion Criterion (IC) used to include relevant studies in our quaternary study was this only one: \textbf{IC1}: The study is a tertiary study in software engineering. Conversely, the Exclusion Criteria (EC) used to exclude the non-relevant studies are: \textbf{(EC1)} The study is not a tertiary study in software engineering; \textbf{(EC2)} The study is written in a language other than English; and \textbf{(EC3)} The full text is not available.

\subsection{Step 2: Conduction}

The conduction took place between September 2021 and January 2022. During this phase, studies were selected and evaluated according to the protocol.

\textbf{Studies Selection:} The automatic search was performed. As a result, 206 potentially relevant studies were selected. We removed duplicated ones and analyzed the remainder. Titles, abstracts, and keywords were read, and IC/EC were applied. The introduction and conclusion sections of each study were also considered and the full text (if necessary). The studies should be tertiary studies and address at least one of the disciplines of SWEBOK \cite{SWEBOK2004}, or those not covered in SWEBOK yet, such as Systems of Systems, or Software Ecosystems, Software Domains, or other important areas for Software Engineering. %Only tertiary studies published in conferences, workshops, symposiums and journals available in the two libraries mentioned above were considered, as they are important and highly relevant sources for Software Engineering. The filtering and selection of these studies were carried out taking into account the study area covered by them. 
As a result of this first selection activity, \textbf{34 tertiary studies} ranging from 2009 to 2021 were included for data extraction.

Most of the studies excluded during the selection step were primary or secondary studies that only mentioned the term `tertiary study', or studies that used the term \textit{tertiary} to denote the third level of formal studies (higher education at the university, in undergraduate courses). %For the primary and secondary studies that answer RQ3 and RQ4, a filter of publication date was added in the search. The criterion used was to carry out the search with the filter from the year following the year of publication of the tertiary study under analysis.
An ID was created for each included study to facilitate its identification and citation. The ID starts with a character referring to the type of publication -- `C' for a conference, `P' for periodical (journal), `W' for a workshop, and `S' for symposium -- followed by the publication year and a unique sequential number from 000 to 034, chronologically ordered, in Appendix A.

\begin{table}[!ht]
\small
\centering
\caption{Data extraction form for selected TS.}
\label{tab:form}
\begin{tabular}{|c|p{7cm}|}
\hline \# & Question\\
\hline 1 & Study title:\\
\hline 2 & Study ID:\\
\hline 3 & Study authors:\\
\hline 4 & Year of publication:\\
\hline 5 & What are the publication years of the most recent and oldest secondary studies included?\\
\hline 6 & Venue (Conference/Journal/Magazine):\\
\hline 7 & What area does the study address? \\
\hline 8 & What subarea does the study address? \\
\hline 9 & Were there already TS in the same area previously published? \\
\hline 10 & Are there secondary studies published after the publication of the included TS? If yes, what is its publication year? \\
\hline {11} & Are there any primary studies published after the publication of the included TS? If yes, what year of publication?\\
\hline
\end{tabular}
\end{table}

\noindent \textbf{Data Extraction.} The form shown in Table \ref{tab:form} was used to support a systematic data extraction from the included TS.
\vspace{-0.2cm}

\subsection{Step 3: Reporting}
\label{sec:algorithm_evaluation}

After the identification and selection of pertinent TS, we examined each study in its entirety to address the posed RQs. Specifically, for addressing research question RQ3, in addition to the thorough examination of the selected tertiary studies, supplementary searches were performed within the ACM Digital Library and IEEE Explorer repositories. These additional searches aimed to identify relevant primary, secondary, and TS published subsequently to the publication of the included TS.
\\\\
\textbf{RQ1: \textit{What are the covered areas in TS?}}

The most discussed area in the included TS was Evidence-Based Software Engineering, with 17 studies (50\% of them) [S2019-008, C2016-011, C2020-013, S2009-015, C2015-018, C2015-022, C2018-023, C202019-024, C2019-025, W2014-026, C2017-027, C2017-028, C2018-029, C2012-030, C2013-032, and C2013-034], followed by Software Construction (7 studies, 20.58\%) [C2020-002, C2012-004, C2011-006, C2017-007, C2012-009, C2012-010, and C2017-033], Software Requirements, with four studies (11.76\%) [C2014-001, C2021-014, C2020-016, and W2018-031], Systems Architecture (three studies, 8.82\%) [C2021-003, C2020-005, and S2020-021] and Software Maintenance with two studies (5.88\%) [C2021-019 and C2017-020].
Other areas addressed in at least one of the studies were Human-Computer Interaction [C2020-017], Software Quality [C2021-019], and IoT [C2018-012]. Therefore, 70\% of the selected tertiary studies addressed evidence-based software engineering or software construction. 

The other subareas covered in only one of the TS are: blockchain and sustainability [C2021-005], blockchain and IoT [C2020-005], software effort [S2017-007], Validity Threat Assessment [C2016-011], Startup Ecosystems [C2020-016], Assistive Technology for Autism [C2020-017], Code Smells [C2021-019], Software Product Line [ C2017-020], Microservices [S2020-021], Quality Assessment [C2015-022], Meta Ethnography [C2019-025], Evidence Distribution and \textit{Pareto's Law} [W2014-026], Research Methodology [C2017 -027], Meta Modeling [C2017-028], Qualitative Research [C2018-029], Requirements Validation [W2018-031] and the Influence of Human Factors on Software Development [S2021-033].\\

%In regards to the subareas, the result was more dispersed in relation to the area, which was wider. The most addressed subareas were gray literature addressed in 4 studies [S2019-008, C2020-013, J2019-024 and S2021-034], evidence-based approach analyzed in 3 studies [C2015-018, C2012-030 and C2013-032] , distributed software development in 2 studies [C2012-004 and C201-009], limitation in search processes in 2 studies [S2009-05 and C2018-023], global software development in 2 studies [C2012-010 and C2021- 014] and agile software development in 2 studies [C2020-002 and C2011-006].

\textbf{RQ2:} \textbf{\textit{What is the year of the oldest and most recent secondary study analyzed by the included tertiary study?}}

The oldest secondary study used in one of the analyzed TS was published in 2004 and the most recent one in 2021. In Table \ref{tab:MarcMNem}, we show the list of the oldest secondary study and the most recent one that was included in each TS analyzed. The aim of this RQ was verifying the time lapse between the publication of the secondary studies considered and the year of publication of the analyzed TS. We observed that, in some of the most recent secondary studies considered by a TS, a time-lapse greater than one year exists. This happens in the studies C2020-002, C2012-004, C2020-013, S2009-015, C2015-018 , C2015-022, C2019-024, C2019-025, W2014-026, C2018-029, W2018-031, C2013-032 and S2021-034. %In analysis and verification of the existence of primary and secondary studies in this period of time, we found that in the case of studies C2020-002, C2013-032 and S2021-034 there are both primary and secondary studies that were not considered by them. And in the case of studies C2012-004, C2020-013, C2019-025, C2018-029, W2018-031 there are primary studies that have been published that may cause some degree of anachronism in these tertiary studies.
A large time lapse between the secondary studies considered by the analyzed TS may be indicative of possible anachronism, as occurred in the case of Cadavid et al. \cite{Cadavid2020}.\\
\vspace{-0.4cm}
\begin{table}[!ht]
\centering
\small
\caption{Years of publication of the oldest and most recent secondary studies considered by each tertiary study.}
\label{tab:MarcMNem} 
\begin{tabular}{|c|p{2cm}|c|p{2cm}|}
\hline Study ID & Time interval of the secondary studies considered in the included TS & Study ID & Time interval of the secondary studies considered in the included TS \\
\hline C2014-001 & 2006-2014 & C2015-018 & 2003-2013\\
\hline C2020-002 & 2013-2018 &  C2021-019 & 2015-2020\\
\hline C2021-003 & 2018-2020 & C2017-020 & 2008-2016\\
\hline C2012-004 & 2006-2010 & S2020-021 & 2016-2019\\
\hline C2020-005 & 2016-2020 & C2015-022 & 2007-2009\\
\hline C2011-006 & 2008-2011 & C2018-023 & 2016-2017\\
\hline C2017-007 & 2006-2016 & C2019-024 & 2004-2010\\
\hline S2019-008 & 2009-2019 & W2014-026 & 2004-2012\\
\hline C2012-009 & 2009-2011 & C2017-027 & 2004-2016\\
\hline C2012-010 & 2005-2011 & C2017-028 & 2005-2016\\
\hline C2016-011 & 2004-2015 & C2017-028 & 2005-2016\\
\hline C2018-012 & 2013-2017 & C2018-029 & 2005-2015\\
\hline C2020-013 & 2004-2012 & C2012-030 & 2004-2011\\
\hline C2021-014 & 2009-2021 & W2018-031 & 2002-2008\\
\hline S2009-015 & 2004-2007 & C2013-032 & 2005-2011\\
\hline C2020-016 & 2016-2019 & S2021-033 & 2018-2020\\
\hline C2020-017 & 2015-2019 & S2021-034 & 2017-2019\\
\hline 
\end{tabular} 
\end{table}

\textbf{RQ3: \textit{Are there secondary studies published after the TS in the same area that might make it anachronic? If yes, what is the year of publication?}}

To verify whether the TS and the secondary studies included by them really dealt with the same area, a complete reading of the TS was carried out and the summary, introduction, and results items of each secondary study surveyed were read. The full report with each case of secondary studies published in the same area (after the publication of a TS that did not consider it) of each TS included and analyzed here can be found in an external link\footnote{\url{bit.ly/3MRHcRd}}, in Portuguese.

We found that the area addressed in at least one of the 34 selected TS was already addressed in a secondary study that was published after the year of publication of the analyzed TS. This occurs in the areas that were addressed by 22 TS: C2014-001, C2020-002, C2012-004, C2020-005, C2011-006, C2014-007, S2019-008, C2012-009, C2012-010, C2016-011, C2012-012, S2009-015, C2020-017, C2015-018, C2017-020, S2020-021, C2015-022, C2017-027, C2018-029, C2012-030, and C2013-032. Therefore, 64.70\% of the TS selected may be out of date due to the existence of secondary studies that addressed the same area but were published after the publication of that TS. Among the analyzed TS that show some degree of anachronism, four were published in 2020, two in 2019, two in 2018, three in 2017, one in 2016, two in 2015, one in 2014, one in 2013, four in 2012, one in 2011 and one in 2009. This reveals that even recent studies were potentially anachronic soon after they were published.

\textbf{RQ4: \textit{Are there primary studies published after the TS in the same area that might make it anachronic? If yes, what is the year of publication?}}

When it comes to primary studies, the same area addressed in at least one of the 34 studies has also been addressed in primary studies that were published after the TS was analyzed. This occurs in the areas that were addressed by 23 studies: C2014-001, C2020-002, C2012-004, C2020-005, C2011-006, C2014-007, S2019-008, C2012-009, C2012-010, C2016-011, C2012-012, C2020-013, S2009-015, C2020-017, C2015-018, C2017-020, S2020-021, C2015-022, C2019-024, C2017-029, C2012-030, and C2013-032. Therefore, 67.64\% of the tertiary studies selected may be out of date due to the existence of primary studies that addressed the same area but were published in a year after the year of publication of the tertiary study. A percentage 2.86\% greater than that was detected in the answer to question RQ3. The area most covered by these tertiary studies is also evidence-based software engineering which was addressed in nine studies, software construction addressed in six studies, systems architecture in two studies, and software requirements in two studies. In addition to these areas, human-computer interaction and software maintenance were addressed in at least one of the studies. Among the analyzed TS that show some degree of anachronism, as there are primary studies that were published in later years but addressing the same area as at least one of the mentioned TS, five were published in 2020, two in 2019, two in 2018, three in 2017, one in 2016, two in 2015, one in 2014, one in 2013,  four in 2012, one in 2011 and one in 2009. An example is the study C2014-001, which addressed mapping studies in requirements engineering and this same area was addressed in the following primary study published after the publication of study C2014-001 \cite{Holm2015}. Similarly, each case is discussed in detail in a full report available in an external link\footnote{\url{bit.ly/3MRHcRd}}, in Portuguese.

\section{Discussion and Implications}\label{sec:progs} 

%Particularly analyzing the responses to the posed RQs, we observe that some areas addressed in the analyzed tertiary studies were also addressed later in both primary and secondary studies. This occurs with the areas that were addressed in 22 of the TS, namely C2014-001, C2020-002, C2012-004, C2020-005, C2011-006, C2014-007, S2019-008, C2012-009, C2012- 010, C2016-011, C2018-012, S2009-015, C2020-017, C2015-018, C2017-020, S2020-021, C2015-022, C2019-024, C2017-027, C2012-030 and C2013-032.

%The areas addressed in these 22 studies are: Software requirements mapping study [C2014-001], distributed software development [C2012-004 and C2012-009], agile software development [C2020-002 and C2011-006], gray literature [S2019-008 and C2019-025], global software development [C2012-010], threats to validity [C2018-012], limitation in search processes [S2009-015], assistive technology for people with autism spectrum [C2020 -017], evidence-based approach [C2015-018, C2012-030 and C2013-032], software product line [C2017-020], microservices [S2020-021], quality assessment [C2015-022] and methodology of research [C2017-027].
Systematic studies take time to be conducted/concluded and the publication process in journals is also time-consuming. Examples include Tore Dybå and Torgeir Dingsøyr \cite{DYBA2008}, published in 2008, but which covers studies up to 2005; and Kamei \cite{Kamei2021b}, with studies included up to 2018, but published in 2021. We conjecture that reviews on theoretic topics maybe can be anachronic more slowly than those associated with technologies. The main findings (\textbf{F$_n$}) of this study are: \textbf{(F$_1$)} 60\% (21 out of) the 34 TS analyzed show evidence that may have some degree of anachronism because of primary and secondary studies published (and not considered in the TS) after their publication. Those novel studies could have presented new results, concepts, and models from those contained in the analyzed TS; %We also analyzed that not only older TS have evidence of outdating, since most of the studies that showed this evidence were published in 2012 [C2012-004, C2012-009, C2012-010 and C2012-030] and 2020 [C2020-002, C2020-005, C2020-017 and S2020-21].
%\textbf{(F$_2$)} 78.26\% of the 34 tertiary studies %[C2014-001, C2020-002, 2012-004, C2017-007, S2019-008, C2012-009, C2012-010, C2018-012, C2020-013, C2020 -017, C2017-020, S2020-021, C2015-022, C2019-024, C2017-027, C2012-030, W2018-031 and 2013-032], mentioned in RQ4, have evidence of some degree of anachronism due to the existence of primary studies addressing the same area published in the year following the year of publication of each analyzed study; 
\textbf{(F$_2$)} Among the 21 TS that show evidence of a greater degree of anachronism because there are both primary and secondary studies in the same area published after the TS publication, in 71.42\% [C2014-001, C2020-002, C2012-004, C2017-007, S2009-008, C2012-009, C2012-010, C2018-012, C2020-017, C2017-020, S2020-022, C2019-024, C2017-027, C2012-030, and C2013-032] of the cases, these primary and secondary studies were published in the next year after the publication of the analyzed TS; and \textbf{(F$_3$)} Most of the TS that did not present evidence of anachronism, as expected, were published in 2021 [C2021-003, C2021-014, C2021 -019, C2021-020, C2021-033, and S2021-035], except for W2014-026, C2017-028, C2018-023, C2019-025, C2020-016. Then, recent TS will be less subject to anachronism, since there was not enough time for the publication of other studies that outdate the TS until the search is executed. Another possible reason is the \textbf{specificity of the topic} covered in the TS. We claim that specific topics tend to not being anachronic fastly due to the scarcity of novel studies (mainly secondary studies). For instance, some of the topics covered among the not anachronic TS involve requirements engineering in startups, meta-ethnography in software engineering, Model-Based Systematic Review, and Pareto's law in software engineering.
\\
\noindent\textbf{The nature of quaternary studies.} %As stated earlier, the secondary \textbf{aim of this research} was \textbf{to gain enhanced insights into the fundamental attributes that may define a quaternary study}. It is clear that 
Researchers broadly know that the type of information that can be obtained when conducting a systematic review (whether it is a secondary, tertiary, or quaternary study) depends on the questions posed and the depth of analysis, in addition to the systematic way in which it was conducted. %However, it can be said that a secondary study, when gathering evidence from primary studies, generally has a narrower scope, potentially providing evidence about something more specific in an area. 
Secondary studies are generally able to gather statistical evidence from an area and allow, for instance, a more accurate decision regarding the effectiveness of a drug to treat a disease (in the case of medicine) or the real impact of using a software development methodology, in the case of SE. %Systematic Mappings usually include only population and \cite{petersen} intervention, so they are more committed to surveying the state of the art on a topic and confirming research opportunities.
When analyzing the TS included, we could observe that \textit{TS offer a broader panorama of their area of investigation}. While secondary studies in SE raise the state of the art on more specific topics (for example, the architectural description activity in SoS in case of \cite{Cadavid2020}), TS generally propose to map an area more broadly, such as Software Construction or Systems Architecture.

In turn, a quaternary study, when investigating TS, goes up one more level. We do not have a conclusive answer about the nature of quaternary studies, since even TS are still scarce in literature and their nature are also still under investigation. We understand that quaternary would be the last level of reviews, since it would probably not be necessary to conduct other quaternary studies, unless the number of tertiary studies in subareas of SE increase so much that makes it possible to exclude some of them and synthesize evidence from that subarea.

%and is able to \textit{map an even broader area of knowledge}, such as SE as a whole. When analyzing TS, it was possible to analyze how the sub-areas have been investigated over the years. Interestingly, we could see that TS have also been used as a \textbf{meta-resource}, i.e, an evidence-based study to analyze the conduction of other evidence-based software engineering studies, investigating aspects such as the use of automatic searches, gray literature, threats to validity and quality assessment.

We could also see that several of them need to be updated since many of the included studies already have more than 10 years of publication, and consider secondary studies not updated from at least six years before it. Although a prior study recommended the 3PDF method to decide whether update a review or not \cite{Mendes2020}, quaternary studies can be an alternative way to assess the need to update TS in particular. It was possible to observe from the quaternary study that there are sub-areas that had not been their TS updated for at least 15 years, evidencing a need and opportunities for research. While primary studies are published all the time, the phenomenon we observe is that many tertiary studies are born anachronic.
\\
\noindent\textbf{Complementing the Guidelines for the Conduction and Update of TS.} A possible solution for TS anachronism, in addition to searches in digital libraries, could be the use of the \textit{snowballing}, both \textit{backarward} and \textit{forward}, to identify new and old primary and secondary studies that have not yet been found. Furthermore, a recommendation for those using TS as a source of knowledge is to verify the quality of the studies\footnote{We also analyzed the quality of the TS included herein, as it can be checked in the provided external link.}, if it presents limitations and internal and external threats to validity. The conduction of Rapid Reviews is also an alternative to complement tertiary studies to check if there were important advances in the field after the publication of the included secondary studies. We also recommend that, before conducting a TS, researchers should analyze the addressed topic: if it updates frequently (a hot topic), they should be aware that the TS will be subject to become obsolete faster.
\\
\noindent\textbf{Threats to Validity.} Some threats to the validity of our conclusions include (i) missing relevant studies (use of only two scientific databases), (ii) potential bias in study selection, (iii) reliability in the conduction and conclusions presented, and (iv) data extraction. The two databases are highly relevant sources in the Software Engineering areas, which reduces the threat; however, a complementary search could also be performed in other databases, such as Scopus\footnote{\url{https://www.elsevier.com/pt-br/solutions/scopus}}, Google Scholar\footnote{\url{https://scholar.google.com/}} and SOL/SBC\footnote{\url{https://sol.sbc.org.br}}. To avoid a biased process, we defined RQs and derived IC and EC. The RQ and criteria are detailed enough to provide an assessment of how reliable is the final set of included studies, reducing the impact of the selection reliability threat. Moreover, the entire material used in the review is also available and can be scrutinized or even replicated. Regarding data extraction, we conducted consensus meetings until full agreements, increasing the reliability of the data extraction. Another threat to the validity is the fact that the search was conducted in January 2022. Then, studies published during the entire year of 2022 and 2023 were not considered here. An update in the search (as future work) could be performed to alleviate this threat.

\section{Final Remarks}
\label{sec:finalremarks}

The main contribution of this vision paper was presenting the results of a review of tertiary studies (aka a quaternary study) in software engineering. We analyzed tertiary studies (TS) in software engineering and the insights that could be provided by such type of analysis (a quaternary study). 206 studies were retrieved and 34 were included for analysis and extraction. The results indicate that 70\% of selected tertiary studies addressed evidence-based software engineering or software construction and 60\% of the TS analyzed in this work have evidence of some degree of anachronism. We also discussed the implications of the findings for the state of practice in the conduction of TS. Our results reveal that, in many cases, there were new studies published after the publication of a TS, which can serve as input to help decide, among the other factors foreseen by the 3PDF approach, whether those SLRs included should be updated or not, or whether even the TS should be updated. We also conclude that more studies and, possibly, guidelines are needed to decide whether to update tertiary studies or not.  %This insight was important to be obtained to settle a better understanding of the status of tertiary studies in the literature of software engineering and, hopefully, support researchers with evidence that point out to the need to refine even more the procedures used to conduct tertiary studies.
Future work includes replicating this study by considering more scientific databases and expanding the concept and maturity of quaternary studies and the proposition of a model of maturity for TS. Also, future investigation should gather evidence on the differences between secondary, tertiary, and quaternary studies and possible changes needed in current guidelines on WHEN and HOW to update systematic studies.

%%
%% The next two lines define the bibliography style to be used, and
%% the bibliography file.
\bibliographystyle{ACM-Reference-Format}
\bibliography{main}
\vspace{-0.2cm}
\appendix
\section{Appendix - Included TS}
{\small
\noindent [S2009-015] The impact of limited search procedures for systematic literature reviews: A participant-observer case study \cite{Kitchenham2009b};\newline
\noindent  [C2011-006] Signs of Agile Trends in Global Software Engineering Research: A Tertiary Study  \cite{BARROSJUSTO2018};\newline
\noindent  [C2012-004] A Systematic Tertiary Study of Communication in Distributed Software Development Projects \cite{Santos2012};\newline
\noindent  [C2012-009] Systematic Literature Reviews in Distributed Software Development: A Tertiary Study  \cite{Marques2021};\newline
\noindent [C2012-010] Systematic literature reviews in global software development: A tertiary study \cite{Verner2012};\newline
\noindent [C2012-030] What scope is there for adopting evidence-informed teaching in SE? \cite{Budgen012};\newline
\noindent [C2013-032] A tertiary study - experiences of conducting systematic literature reviews in software engineering \cite{Imtiaz2013};\newline
\noindent [C2014-001] Systematic Reviews in Requirements Engineering: A Tertiary Study  \cite{Bano2014};\newline
\noindent [W2014-026] Does Pareto's law apply to evidence distribution in software engineering? an initial report \cite{Tang2014};\newline
\noindent [C2015-018] Trends and perceptions of evidence-based software engineering research in Malaysia \cite{Salleh2014};\newline
\noindent [C2015-022] Quality assessment of systematic reviews in software engineering: a tertiary study \cite{Zhou2015}; \newline
\noindent [C2016-011] A Map of Threats to Validity of Systematic Literature Reviews in Software Engineering \cite{Zhou2016}; \newline
\noindent [C2017-007] Consolidating evidence based studies in software cost effort estimation: A tertiary study \cite{Pillai2017};\newline
\noindent [C2017-020] Systematic Studies in Software Product Lines: A Tertiary Study \cite{Marimuthu2017};\newline
\noindent [C2017-027] Experience-based guidelines for effective and efficient data extraction in systematic reviews in software engineering \cite{Garousi2017};\newline
\noindent [C2017-028] A Model-Based Approach to Systematic Review of Research Literature \cite{Barat2017};\newline
\noindent [C2018-012] AAL Platforms Challenges in IoT Era: A Tertiary Study \cite{Duarte2018};\newline
\noindent [C2018-023] How do Secondary Studies in Software Engineering report Automated Searches? A Preliminary Analysis \cite{Singh2018};\newline
\noindent [C2018-029] Synthesizing qualitative research in software engineering: a critical review \cite{Huang2018};\newline
\noindent [W2018-031] Experiences in using practitioner's checklists to evaluate the industrial relevance of requirements engineering experiments \cite{Daneva2018};\newline
\noindent [S2019-008] Multivocal literature reviews in software engineering: Preliminary findings from a tertiary study  \cite{Neto2019};\newline
\noindent [C2019-024] The Use of Grey Literature Review as Evidence for Practitioners \cite{Kamei2019};\newline
\noindent [C2019-025] A Review of Meta-ethnographies in Software \cite{Fu2019};\newline
\noindent [C2020-002] Bibliometric Analysis of the Tertiary Study on Agile Software Development using Social Network Analysis \cite{Bayram2020};\newline
\noindent [C2020-005] Blockchain-based Solutions for IoT: A Tertiary Study\cite{XU2020};\newline
\noindent [C2020-013] On Using Grey Literature and Google Scholar in Systematic Literature Reviews in Software Engineering \cite{Yasin2020};\newline
\noindent [C2020-016] On the Pragmatics of Requirements Engineering Practices in a Startup Ecosystem \cite{Alves2020};\newline
\noindent [C2020-017] Scrutinizing Reviews on Computer Science Technologies for Autism: Issues and Challenges \cite{Epifanio2020};\newline
\noindent [S2020-021] Microservice Architecture: A Tertiary Study \cite{Costa2020};\newline
\noindent [C2021-003] Blockchain and Sustainability: A Tertiary  \cite{Jiang2021};\newline
\noindent [C2021-014] A Study on Mitigating the Communication Coordination Challenges During Requirement Change Management in Global Software Development \cite{Qureshi2021};\newline
\noindent [C2021-019] Tertiary Study on landscaping the review in code smells \cite{Yaqoob2021};\newline
\noindent [S2021-033] Human Factors and their Influence on Software Development Teams: A Tertiary Study \cite{Dutra2021};\newline
\noindent [S2021-034] What Evidence We Would Miss If We Do Not Use Grey Literature? \cite{Kamei2021};
}

\end{document}